\documentclass[fleqn,usenatbib]{mnras}

\usepackage{newtxtext,newtxmath}

\usepackage[T1]{fontenc}

\DeclareRobustCommand{\VAN}[3]{#2}
\let\VANthebibliography\thebibliography
\def\thebibliography{\DeclareRobustCommand{\VAN}[3]{##3}\VANthebibliography}

\usepackage{graphicx}	
\usepackage{amsmath}	

\title[The nature of {\it tilted} supercritical accretion discs]{The nature of {\it tilted} supercritical accretion discs}

\author[P. C. Fragile et al.]{
P. Chris Fragile,$^{1,2}$\thanks{E-mail: fragilep@cofc.edu (PCF)}
Matthew J. Middleton,$^{3}$
Brooks Brasseur,$^{1}$
Deepika A. Bollimpalli$^{4}$
and Zach Smith$^{1}$\\
$^{1}$Department of Physics and Astronomy, College of Charleston, 66 George Street, Charleston, SC 29424, USA\\
$^{2}$Center for Computational Astrophysics, Flatiron Institute, 162 5th Avenue, New York, NY 10010, USA\\
$^{3}$Department of Physics and Astronomy, University of Southampton, Highfield, Southampton SO17 1BJ, UK\\
$^{4}$Department of Astronomy, Astrophysics \& Space Engineering, Indian Institute of Technology Indore, Simrol, Indore 453552, Madhya Pradesh, India
}

\date{Accepted XXX. Received YYY; in original form ZZZ}

\pubyear{2025}

\begin{document}
\label{firstpage}
\pagerange{\pageref{firstpage}--\pageref{lastpage}}
\maketitle

\begin{abstract}
In this paper, we report on the first 3D general relativistic radiation magnetohydrodynamic simulations of large supercritical accretion discs that are tilted with respect to the black hole spin axis. We explore a range of black hole spin parameters (from $a_* = -0.9$ to 0.9), initial tilts (in the range from $\beta_0 = 0^\circ$ to $30^\circ$), and target mass accretion rates. We first confirm that, for all the untilted simulations, the Eddington accretion limit is obeyed ($\dot{M}_\mathrm{BH} \lesssim \dot{M}_\mathrm{Edd}$), consistent with our previous findings. However, for tilted discs we find that the mass accretion rate can be enhanced by up to a factor of ten and that factor depends linearly on tilt $\dot{M}_\mathrm{BH} \propto \beta_0 \ge \dot{M}_\mathrm{Edd}$. This could be an important aspect in solving the puzzle of the growth of the first supermassive black holes. We also find that for a given tilt, the mass accretion rate enhancement is proportional to the magnitude of the spin. Additionally, we find that tilted supercritical accretion discs are more advective than their untilted counterparts. We attribute all of these differences to the presence of standing shocks in the inner regions of the accretion flow, a feature unique to tilted discs. 
\end{abstract}




\section{Introduction}
\label{sec:introduction}

Supercritical accretion, where matter is fed into a black hole of mass $M$ at a rate above the nominal Eddington limit, $\dot{M}_\mathrm{Edd} \equiv L_\mathrm{Edd}/\eta c^2$, where $L_\mathrm{Edd} = 1.3 \times 10^{38} (M/M_\odot) ~\mathrm{erg~s}^{-1}$ is the Eddington luminosity and $\eta$ is the radiative efficiency, is of primary importance for the appearance of X-ray bright systems such as ultra-luminous X-ray sources \citep[ULXs;][]{King01, Kaaret17, King23}, transients such as tidal disruption events \citep[TDEs;][]{Dai18, Wu18}, and the extremely rapid growth of some supermassive black holes (SMBHs) at very high redshifts \citep{Volonteri05, Schneider23, Taylor25}. All previous numerical work on supercritical accretion \citep[e.g.,][]{Ohsuga05, Jiang14, Sadowski16, Takahashi18, Asahina22, Utsumi22, Fragile25, Zhang25}, except \citet{Asahina24}, has assumed that the spin axis of the black hole is aligned with the angular momentum axis of the accreting gas. In our parlance, such discs are ``untilted'' or ``aligned.'' 

However, this may not be the norm in nature. For example, in the cases of TDEs and SMBHs, there is no way for the accretion flow at large scales to know about or be affected by the orientation of the spin axis of the black hole; therefore, there is no reason to expect the accretion disc to be aligned initially \citep{King06}. In fact, in such cases, retrograde accretion, where the angular momentum vector of the accreting gas points more than $90^\circ$ away from the spin axis of the black hole, may be just as common as prograde \citep{King08}. Even in the case of ULXs, misalignment may be fairly common, though likely less extreme than for TDEs and SMBHs. For ULXs, the orientation of the outer disc is fixed by the binary orbit, whereas the spin orientation of the black hole is set at its birth. Thus, if the supernova explosion that created the black hole were asymmetric, as it appears many are \citep[e.g.,][]{Grefenstette14, Boggs15}, then the black hole spin axis could be misaligned with respect to the binary (and hence the disc) even if the system were aligned prior to the supernova \citep{Fragos10}.

The resulting ``tilted'' discs, where the angular momenta are inclined with respect to the black hole spin axes, are subject to Lense-Thirring precession, just like particles in tilted orbits. For particles, the strong radial dependence of Lense-Thirring precession, $\Omega_\mathrm{LT}(r) \approx 2 a_* (G M)^2/(c r)^3$, means that those on larger orbits would precess much more slowly than those on smaller orbits. In discs, the strong radial dependence would naively cause them to twist and warp. The key is in how this warping is transmitted through the disc. For thin discs, it is diffused outward via viscous stresses, allowing the inner disc to settle into the symmetry plane of the black hole \citep{Bardeen75}. For compact thick discs, it is possible for the precession to be communicated across the entire disc fast enough for it to precess as a global structure \citep{Fragile07, Liska18, White19}. This was predicted for supercritical (but still compact) tilted accretion discs by \citealt{Middleton18, Middleton19} and later confirmed by \citet{Asahina24}.

Following their formation, there is a tendency for tilted accretion discs to align with the black hole due to their mutual tidal interactions \citep{Rees78}. However, for black hole ULXs and TDEs, the alignment timescale is longer than their observational lifetimes \citep{Scheuer96, King16}, and for the growth of SMBHs, each new accretion episode has the possibility of reorienting the fuel supply, potentially resulting in repeated tilted configurations \citep{Kinney00, Middleton16}. 

What remains uncertain from previous work is into which regime large, Keplerian supercritical discs fall. Could they even exhibit their own unique behavior? This question motivates our present work in which we extend our previous study of large supercritical accretion discs \citep{Fragile25} by considering tilted cases. We explore how tilt affects the mass accretion rate, luminosity, and radiative efficiency of tilted discs. We find that an important feature of tilted supercritical discs is a pair of standing shocks that form close to the inner edge of the disc. These shocks act to extract angular momentum, shortening the time that matter orbits in the inner disc and, therefore, giving it less time to radiate.

We report all these results throughout the remainder of this paper. In Section \ref {sec:setup}, we review the details of our numerical setup, specifically the parameters we explore and how the tilted simulations are initialized. In Section \ref{sec:results}, we present our results comparing untilted and tilted simulations, focusing primarily on those aspects that are unique to tilted discs. We also include dedicated sections on two of the most prominent features associated with tilted discs: standing shocks (covered in Section \ref{sec:standing_shocks}) and Lense-Thirring precession (Section \ref{sec:precession}). Finally, we end in Section \ref{sec:conclusions} with some discussion and concluding thoughts. 

Throughout most of this work, whenever we report a mass accretion rate, we report it as $\dot{m} = \dot{M}/\dot{M}_\mathrm{Edd}$. In other words, we scale our mass accretion rates to Eddington assuming the nominal Novikov-Thorne efficiency $\eta_\mathrm{NT}$ for the corresponding black hole spin.

\section{Numerical Setup}
\label{sec:setup}

All of the simulations in this work are performed using the general relativistic radiation MHD (GRRMHD) code Cosmos++ \citep{Anninos05, Fragile14}. The detailed setup of our simulations follows the same procedure as described in Section 2 of \citet{Fragile25}. As in that work, all our untilted simulations are initialized from the \citet{Novikov73} generalization of the Shakura-Sunyaev \citep{Shakura73} thin disc with viscosity parameter $\alpha_\mathrm{SS} = 0.02$ and a nominal, target mass accretion rate of $\dot{m}_0$ (provided in Table \ref{tab:models}). Since our target $\dot{m}_0$ are all $>1$, they should correlate to a critical radius in the disc of $r_\mathrm{cr} \approx 9/4 \dot{m}_0 r_\mathrm{ms}$ \citep{Fukue04, Poutanen07}, where $r_\mathrm{ms}$ is the marginally stable circular orbit radius, which depends on spin. We use the same quadrupole initial magnetic field configuration as in \citet{Fragile25}. All of the simulations use $M = 6.62 M_\odot$, though we would expect our results to apply for any stellar mass black hole. We expand on the $a_* = 0.9$ results of \citet{Fragile25} by considering additional spins of $a_* = -0.9$ (retrograde), 0, and 0.5.

For the tilted disc simulations that are the main focus of this work, we start each one from an untilted simulation that has the same black hole spin and has run for at least $50\,000\,t_g$, where $t_g = GM/c^3$. This provides sufficient time for the original disc to have settled into a quasi-steady state and establish inflow equilibrium out to $r_\mathrm{eq} \gtrsim 30\,r_g$, where $r_g = GM/c^2$. We then introduce the tilt by rotating the black hole angular momentum vector counterclockwise about the $+y$-axis. The result is a black hole spin axis that is tilted away from the $+z$-axis toward the $-x$-axis by an angle $\beta_0$\footnote{For the tilted retrograde case ($a_* = -0.9$), we tilt the black hole spin axis by an angle $\beta_0$ away from the $-z$-axis toward the $+x$-axis.}. In practice, this is handled by transforming the standard Kerr metric using a simple rotation as described in \citet{Fragile05}. This choice to tilt the black hole instead of the disc allows us to maintain similar resolution over most of the disc in all cases. It also facilitates our ability to easily restart the tilted simulations from already evolved untilted ones. In this work, we explore tilts of $\beta_0 = 7.5^\circ$, $15^\circ$, $22.5^\circ$, and $30^\circ$. We evolve each tilted simulation for a total duration $t_\mathrm{dur}$ of at least $22\,000\,t_g$ beyond the end of the original, untilted simulation to allow it to reach a new quasi-steady state. 

Table \ref{tab:models} provides details of all of the simulations presented in this paper, with the model naming convention being ``a$X$'' for the spin, ``r$X$'' for the target critical radius (related to the nominal target $\dot{m}_0$), and ``b$X$'' providing the tilt \footnote{If there is no b$X$ in the name, then that implies an untilted ($\beta_0 = 0^\circ$) simulation, consistent with our naming convention in \citet{Fragile25}.}; $h$ parametrizes the concentrated polar coordinate, $\theta = x_2 + h \sin(2 x_2)$, that we use to better resolve the disc region.

\begin{table*}
\centering
\begin{tabular}{lcccccccccccc}
\hline
\hline
& $a_*$ & $\eta_\mathrm{NT}$ & $\dot{m}_0$ & $\beta_0\,(^\circ)$ & $h$ & $t_\mathrm{dur}/t_g$ & $r_\mathrm{eq}/r_g$ & $\langle\dot{m}_\mathrm{in}(r_\mathrm{eq})\rangle_t$ & $\langle\dot{m}_\mathrm{BH}\rangle_t$ & $\frac{\langle L_\mathrm{out}(r_\mathrm{eq})\rangle_t}{L_\mathrm{Edd}}$ & $\frac{\langle L_\mathrm{kin}(r_\mathrm{eq})\rangle_t}{L_\mathrm{Edd}}$ & $\langle\eta\rangle_t$ \\
\hline
a-9r200 & -0.9 & 0.039 & 10 & 0 & 0.15 & 100 000 & 25 & 6.9 & 2.3 & $\le 2.3$ & $\le 2.8$ & $\le 0.04$ \\
a-9r200b15 & -0.9 & 0.039 & 10 & 15 & 0.15 & 23 167 & 43 & 27 & 5.3 & $\le 4.7$ & $\le 6.9$ & $\le 0.03$ \\
a0r100 & 0 & 0.057 & 7 & $\cdots$ & 0.25 & 51 042 & 30 & 11 & 2.3 & $\le 3.3$ & $\le 2.2$ & $\le 0.08$ \\
a5r50 & 0.5 & 0.082 & 5 & 0 & 0.35 & 76 424 & 22 & 16 & 1.7 & $\le 3.7$ & $\le 3.3$ & $\le 0.2$ \\
a5r50b15 & 0.5 & 0.082 & 5 & 15 & 0.35 & 24 855 & 22 & 19 & 2.4 & $\le 4.2$ & $\le 4.1$ & $\le 0.1$ \\
a9r20 & 0.9 & 0.16 & 4 & 0 & 0.35 & 150 000 & 49 & 42 & 1.2 & $\le 5.0$ & $\le 2.6$ & $\le 0.7$ \\
a9r20b7.5 & 0.9 & 0.16 & 4 & 7.5 & 0.35 & 22 113 & 21 & 30 & 2.9 & $\le 3.0$ & $\le 3.6$ & $\le 0.2$ \\
a9r20b15 & 0.9 & 0.16 & 4 & 15 & 0.35 & 25 000 & 60 & 73 & 4.1 & $\le 5.4$ & $\le 3.6$ & $\le 0.2$ \\
a9r20b22.5 & 0.9 & 0.16 & 4 & 22.5 & 0.35 & 22 957 & 85 & 130 & 6.6 & $\le 7.5$ & $\le 4.9$ & $\le 0.2$ \\
a9r20b30 & 0.9 & 0.16 & 4 & 30 & 0.35 & 25 000 & 85 & 140 & 6.9 & $\le 8.2$ & $\le 5.2$ & $\le 0.2$ \\
\hline
\hline
\end{tabular}
\caption{Simulation models and parameters. See text for meaning of symbols. The output values for model a9r20 in this table differ from those provided in \citet{Fragile25} because here we time-average over a different period of the simulation.}
\label{tab:models}
\end{table*}

\section{Results}
\label{sec:results}

In this section, we systematically review the results of our simulations. Most of the section focuses on results unique to tilted supercritical accretion discs, though we start by expanding on one of our main conclusions from \citet{Fragile25}.

\subsection{Untilted discs (still) obey the Eddington limit}

In \citet{Fragile25}, we found that large, Keplerian supercritical accretion discs that are aligned with the black hole spin axis settle close to a net mass accretion rate of $\dot{m}_\mathrm{net} = \dot{m}_\mathrm{in}-\dot{m}_\mathrm{out} \approx 1$ over the radii where the simulations reach equilibrium, with
\begin{equation}
\dot{M}_\mathrm{in}(r,t) = -\int \int \sqrt{-g} \rho \mathrm{min}\{u^r,0\} {\rm d}\theta {\rm d} \phi
\end{equation}
being the inflowing mass accretion rate and
\begin{equation}
\dot{M}_\mathrm{out}(r,t) = \int \int \sqrt{-g} \rho \mathrm{max}\{u^r,0\} {\rm d}\theta {\rm d} \phi 
\end{equation}
the outflowing rate. In other words, these simulated untilted discs closely obey the Eddington limit ($\dot{M}_\mathrm{BH} \approx \dot{M}_\mathrm{Edd}$). This is true even though the inward mass flux (measured at large radii), $\dot{m}_\mathrm{in} (r>100r_g)$, can exceed $1\,000$ in some of our simulations\footnote{The inward mass accretion rates can exceed our reported values for $\dot{m}_0$ by such large amounts because in these simulations $\dot{m}_\mathrm{in}$ accounts for all inward movement of gas, including that associated with waves and convection, not just accretion.}. As was shown in \citet{Fragile25}, this throttling of the mass accretion is possible because the outflowing mass flux $\dot{m}_\mathrm{out}(r)$ at each radius adjusts itself to very nearly cancel $\dot{m}_\mathrm{in}(r)$, so that at all radii $\dot{M}_\mathrm{net} (r) \approx \dot{M}_\mathrm{Edd}$, consistent with the classical critical disc picture \citep{Shakura73, Fukue04}. 

\begin{figure}
\centering
\includegraphics[width=1.0\linewidth,trim=0mm 0mm 0mm 0,clip]{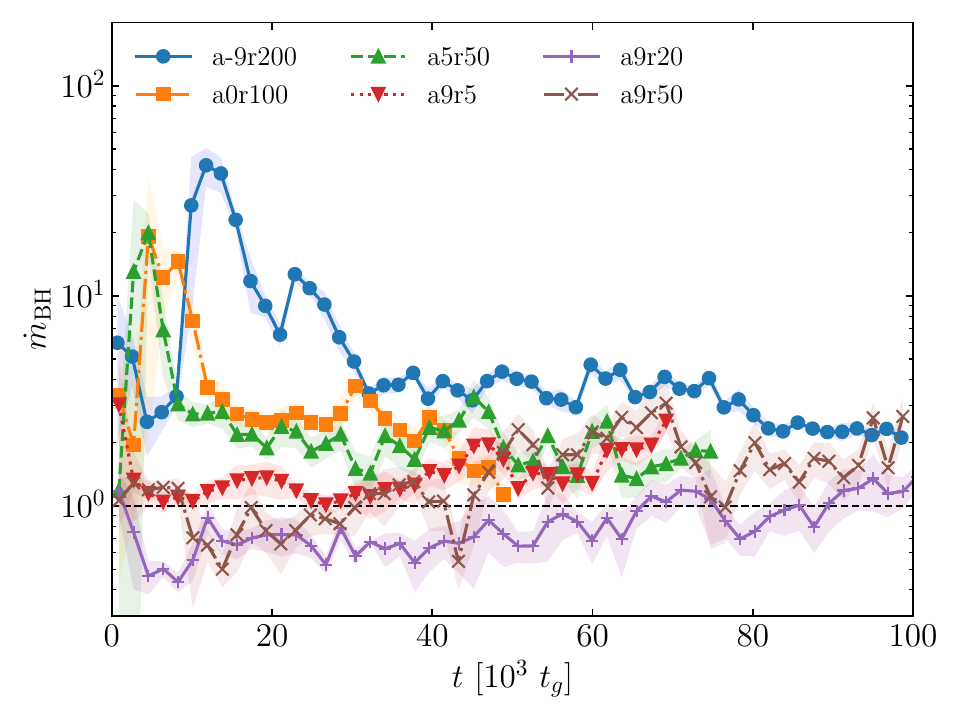}
\caption{Mass accretion rate through the black hole event horizon for our untilted ($\beta_0=0^\circ$) simulations in units of the Eddington accretion rate $\dot{m}_\mathrm{BH} = \dot{M}_\mathrm{BH}/\dot{M}_\mathrm{Edd}$ (using $\eta_\mathrm{NT}$ to define $\dot{M}_\mathrm{Edd}$), smoothed using averages over 20 consecutive dumps ($\approx 1850\,t_g$ in time). The shaded regions show the $1\sigma$ standard deviations, and the black dashed line shows the Eddington limit.}
\label{fig:mdot_untilted}
\end{figure}

For the expanded set of parameters explored in the current work, we find this limitation of $\dot{M}_\mathrm{BH} \lesssim \dot{M}_\mathrm{Edd}$ continues to hold true for all of the untilted simulations we have tried\footnote{It remains to be proven whether the Eddington limit could still restore itself if these simulations were, say, arbitrarily rescaled to larger densities.}, now exploring multiple values of black hole spin. This is shown in Figure \ref{fig:mdot_untilted}, where we plot the mass accretion onto the black hole as a function of time for six untilted simulations. Note that by the end of each simulation, $\dot{m}_\mathrm{BH}$ has converged to within a factor of 2 of unity. Some cases take longer than others to reach this limit, but once they get there, all the simulations stay within a factor of 2-3. Table \ref{tab:models} reports the mass accretion rates onto the black hole $\langle\dot{m}_\mathrm{BH}\rangle_t$, time-averaged over the final $20\,000\, t_g$ of each simulation.  

If mass accretion is truly limited to $\dot{M}_\mathrm{Edd}$, then this leads to difficulties when trying to grow the first supermassive black holes in the Universe \citep{Smith19}. If one starts from a stellar-mass progenitor ($\lesssim 100 M_\odot$), then one cannot reach the required masses ($\gtrsim 10^9 M_\odot$) in the available time \citep[$\lesssim 700$ Myr;][]{Banados18, Yang21, Taylor25} unless the radiative efficiency somehow remains very low. If the observations are accurate \citep[see, however,][for arguments they may not be]{King24}, then there are only two possible explanations: either the progenitors are larger than expected \citep[see][for ideas in this vein]{Portegies04, Begelman06, Dijkstra08, Mayer10} or the growth is not actually limited to $\dot{M}_\mathrm{Edd}$. As we show in the next section, our tilted simulations support the latter possibility.

\subsection{Tilted discs can exceed the Eddington limit}
\label{sec:mdot_tilted}

\begin{figure}
\centering
\includegraphics[width=1.0\linewidth,trim=0mm 0mm 0mm 0,clip]{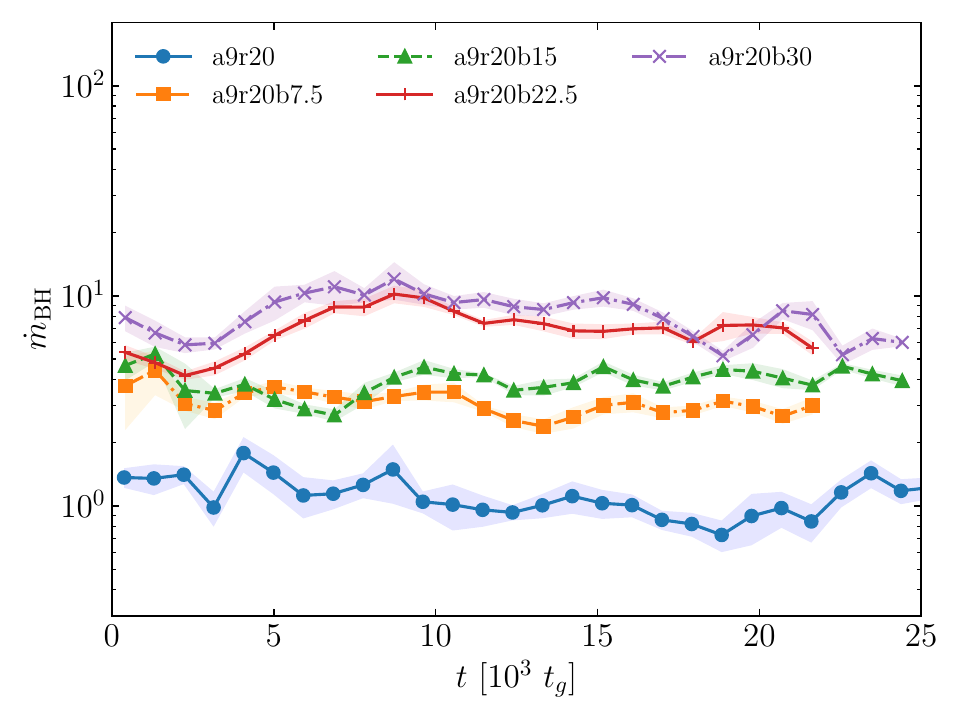}
\caption{Same as Figure \ref{fig:mdot_untilted}, but only for the $a_* = 0.9$ simulations with tilts varying from $\beta_0 = 0^\circ$ to $30^\circ$. For the a9r20 simulation we show the accretion rate starting at $100\,000\,t_g$ (end point of Figure \ref{fig:mdot_untilted}).}
\label{fig:mdot_tilted}
\end{figure}

The conclusion accretion is limited to Eddington does not appear to apply to our tilted disc simulations. In Figure \ref{fig:mdot_tilted}, we present the same time history plot of accretion onto the black hole as in Figure \ref{fig:mdot_untilted}, except now only for our $a_* = 0.9$ simulations with tilts ranging from $\beta_0 = 0^\circ$ to $30^\circ$. There is clearly a systematic trend, where higher values of $\beta_0$ yield higher rates of mass accretion onto the black hole $\dot{m}_\mathrm{BH}$. Since the untilted ($\beta_0 = 0^\circ$) simulation is Eddington-limited, this directly implies that our tilted ($\beta_0 > 0^\circ$) accretion discs exceed this limit and accrete more efficiently. We see from Figure \ref{fig:mdot_tilted} that, for large enough tilts, the mass accretion rate can be as much as five times higher or more.

\begin{figure}
\centering
\includegraphics[width=1.0\linewidth,trim=0mm 0mm 0mm 0,clip]{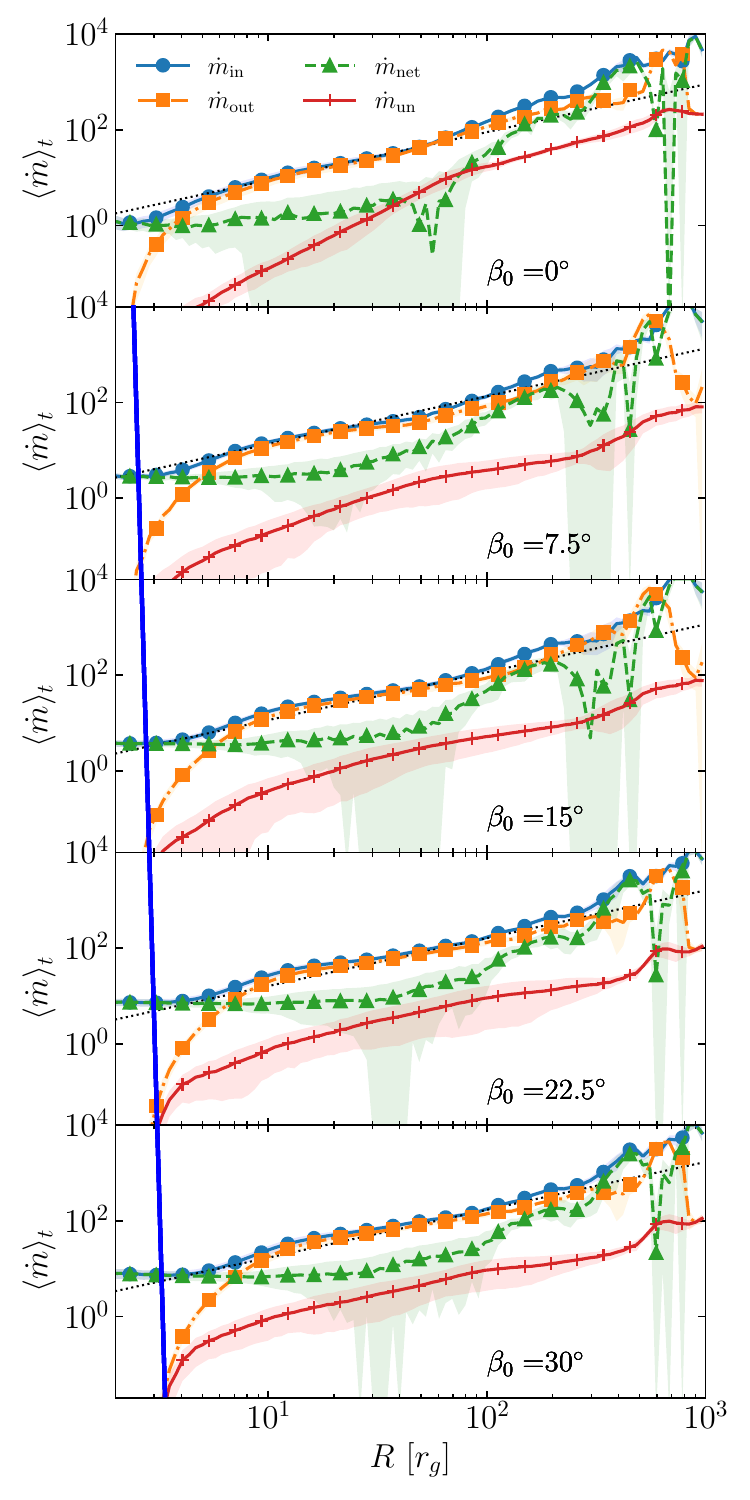}
\caption{Mass fluxes, both inward ($\dot{m}_\mathrm{in}$) and outward ($\dot{m}_\mathrm{out}$), as well as the net $\dot{m}_\mathrm{net} = \dot{m}_\mathrm{in} - \dot{m}_\mathrm{out}$, all scaled to Eddington and time-averaged over the final $20\,000\,t_g$ of each simulation for the a9r20 ({\em top}), a9r20b7.5 ({\em second}), a9r20b15 ({\em third}), a9r20b22.5 ({\em fourth}), and a9r20b30 ({\em bottom}) simulations. The other curves report the portion of $\dot{m}_\mathrm{out}$ that has a positive Bernoulli parameter ($\dot{m}_\mathrm{un}$), plus an analytic estimate for $\dot{m}_\mathrm{in}(r) = \dot{m}_\mathrm{in}(r_\mathrm{cr})r/r_\mathrm{cr}$ (black, dotted curve). The shaded regions show $1\sigma$ standard deviations. The thick blue line connects the minimum radius for which $\dot{m}_\mathrm{out} > 0$ for each simulation, highlighting the fact that the outflow is suppressed as the tilt increases.}
\label{fig:mdot_flux}
\end{figure}

The enhanced accretion rate when the disc is tilted is triggered from very close to the black hole, where we notice the largest deviations in the mass flux profiles. Figure \ref{fig:mdot_flux} shows time-averaged radial profiles of mass flux, including inward, outward, and net, for all the $a_* = 0.9$ simulations. These data are time-averaged over the final $20\,000\,t_g$ of each simulation. Figure \ref{fig:mdot_flux} also includes $\dot{M}_\mathrm{un}$, which represents that portion of $\dot{M}_\mathrm{out}$ that has a positive Bernoulli parameter $Be = -(T^t_t + R^t_t + \rho u^t) > 0$ \citep{Sadowski16} and thus is likely to be unbound and ultimately escape to infinity. 

An important takeaway from Figure \ref{fig:mdot_flux} is that $\dot{m}_\mathrm{out}$ first becomes non-zero increasingly further away from the black hole as the tilt grows (following the thick blue line). Because the outflow is not as efficient at small radii for tilted discs, more matter is able to fall into the hole. Even though this seemingly small difference is most noticeable close to the black hole, it has consequences throughout the disc. This is because the outflow at any radius is an accumulation of all the outflow from smaller radii plus whatever is driven from that radius. A smaller outflow at the innermost edge of tilted discs therefore affects all radii, leading to $\dot{m}_\mathrm{out}(r)$ being lower than it is for the corresponding untilted disc. Since $\dot{m}_\mathrm{out}(r)$ is lower, $\dot{m}_\mathrm{net}$ is consequently higher (since $\dot{m}_\mathrm{in}$ is relatively unchanged). This is apparent from a careful inspection of the profiles in Figure \ref{fig:mdot_flux}.

A similar enhancement of mass accretion was noted in earlier (non-radiative) tilted accretion disc simulations as far back as \citet{Fragile07} (see their Fig. 10a). As we show in Section \ref{sec:standing_shocks}, the enhancement can even be attributed to a similar cause -- standing shocks associated with tilted discs.

\subsection{Radiative luminosity}

\begin{figure}
\centering
\includegraphics[width=1.0\linewidth,trim=0mm 0mm 0mm 0,clip]{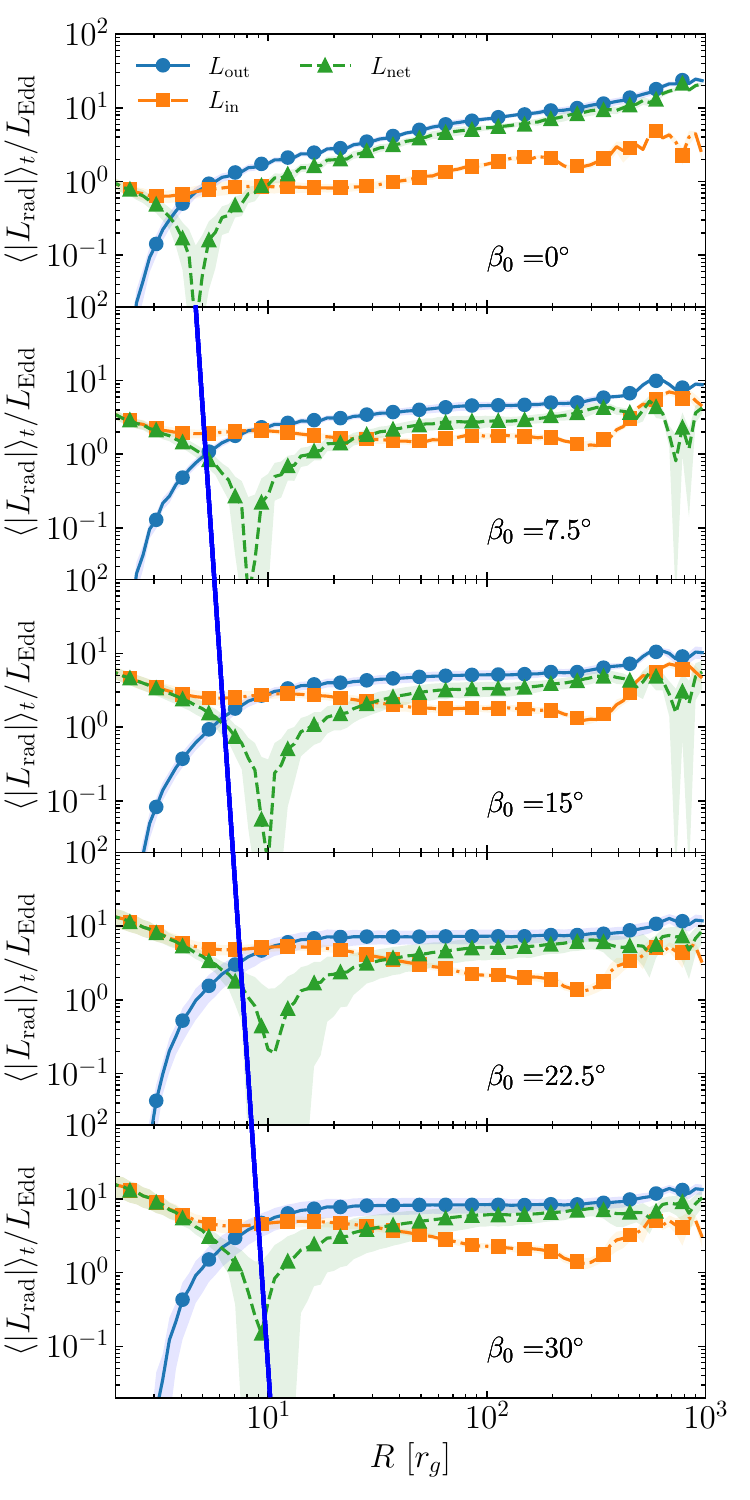}
\caption{Radiative luminosity, both outward ($L_\mathrm{out}$) and inward ($L_\mathrm{in}$), as well as the net $L_\mathrm{net} = L_\mathrm{out} - L_\mathrm{in}$, all scaled to Eddington and time averaged over the final $20\,000\,t_g$ of each simulation for the a9r20 ({\em top}), a9r20b7.5 ({\em second}), a9r20b15 ({\em third}), a9r20b22.5 ({\em fourth}), and a9r20b30 ({\em bottom}) simulations. The shaded regions show $1\sigma$ standard deviations. The trapping radius $r_\mathrm{tr}$ is apparent as the sharp dip in $L_\mathrm{net}$ around $r \lesssim 10 r_g$, where it actually changes sign from inflowing (for $r<r_\mathrm{tr}$) to outflowing (for $r>r_\mathrm{tr}$).}
\label{fig:Lrad_flux}
\end{figure}

Another difference associated with tilted accretion discs is that their trapping radii are located further from the black hole. We demonstrate this in Figure \ref{fig:Lrad_flux}, where we plot the time-averaged radial profiles of the radiative luminosity
\begin{equation}
L_\mathrm{rad}(r,t) = -\int \int \sqrt{-g} R^r_t {\rm d}\theta {\rm d} \phi ~,
\end{equation}
integrated over the full $4\pi$ steradians (complete radial shells). We report both the outward ($u^r_R > 0$) and inward ($u^r_R < 0$) contributions\footnote{Inward luminosity is mostly attributed to photons that are trapped within the optically thick accreting gas.}, as well as the net luminosity, $L_\mathrm{net} = L_\mathrm{out} - L_\mathrm{in}$.   

The point where $L_\mathrm{net}$ changes sign (identified by the sharp dip in the green curves) in Figure \ref{fig:Lrad_flux} is where we identify the trapping radius $r_\mathrm{tr}$. Inside this radius ($r < r_\mathrm{tr}$) most of the radiation is flowing toward the black hole, while outside it ($r>r_\mathrm{tr}$), most of the radiation moves away from the hole. Notice that, similar to how in Figure \ref{fig:mdot_flux} the mass outflow only becomes non-zero further from the black hole when the disc is tilted, the trapping radius in Figure \ref{fig:Lrad_flux} is located further out for tilted discs.

Another way to see this point is to compare the values of $L_\mathrm{in}$ at the event horizon in each case in Figure \ref{fig:Lrad_flux}. Notice how its value increases with increasing tilt. Put a different way, tilted discs are more advective (i.e., deposit more of their radiative energy into the black hole) than the equivalent untilted disc.

\subsection{Mass accretion rate is proportional to tilt and spin}

Careful inspection of Figure \ref{fig:mdot_tilted} shows that not only is the mass accretion more efficient for tilted simulations, but $\dot{m}_\mathrm{BH}$ appears to scale with the tilt $\beta_0$. In this section we quantify that result and generalize it to other spins. To facilitate this task, we define an accretion ``enhancement'' factor 
\begin{equation}
\xi \equiv \frac{\langle\dot{m}_\mathrm{BH}(\beta_0)\rangle_t}{\langle\dot{m}_\mathrm{BH}(0^\circ)\rangle_t} ~,
\end{equation}
which is the mass accretion rate onto the black hole for tilt $\beta_0$, normalized by the mass accretion rate measured for the corresponding untilted simulation. Using this dimensionless enhancement factor, we can more easily compare different spins and tilts.

\begin{figure}
\centering
\includegraphics[width=1.0\linewidth,trim=0mm 0mm 0mm 0,clip]{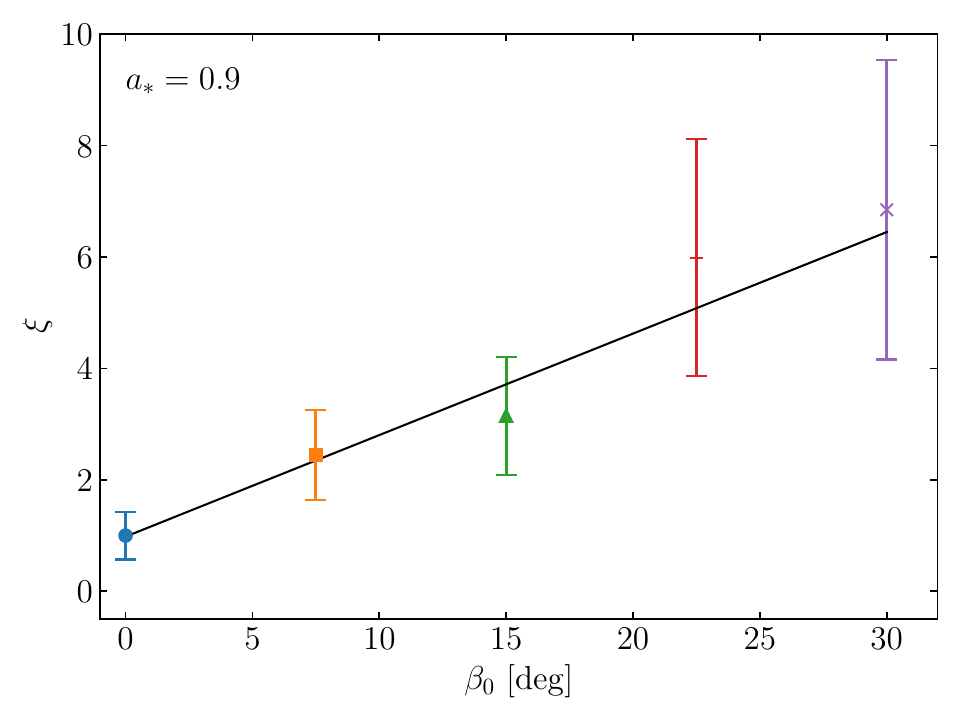}
\includegraphics[width=1.0\linewidth,trim=0mm 0mm 0mm 0,clip]{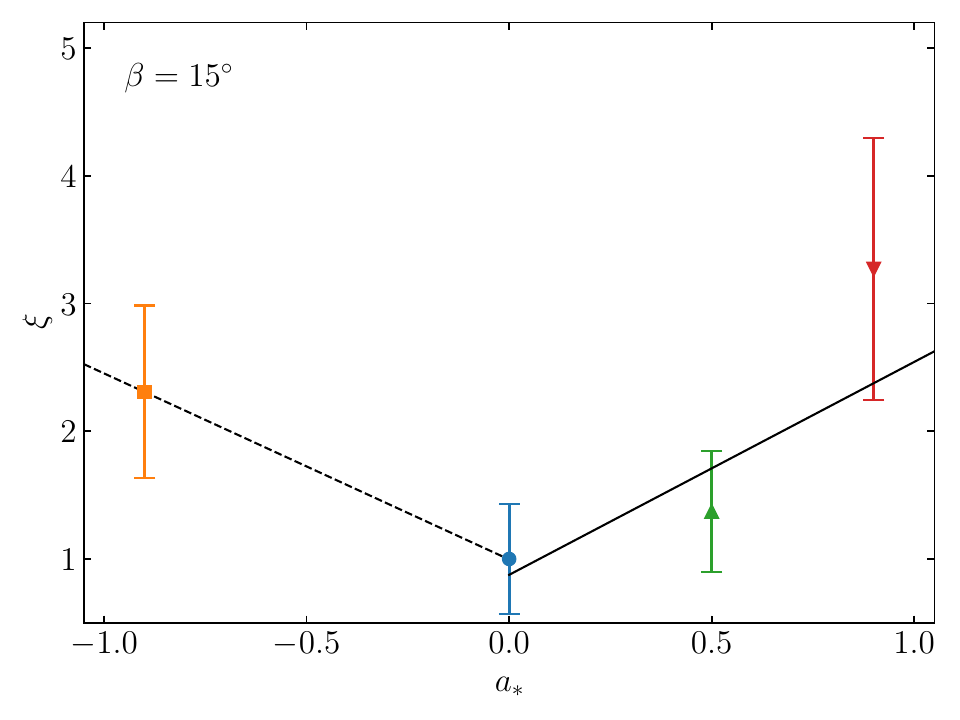}
\caption{{\em Top panel:} Mass accretion enhancement $\xi = \langle\dot{m}_\mathrm{BH}(\beta_0)\rangle_t/\langle\dot{m}_\mathrm{BH}(0^\circ)\rangle_t$ as a function of tilt for our $a_* = 0.9$ simulations. Values are extracted by time-averaging over the final $20\,000\,t_g$ of each simulation. The error bars represent the $1\sigma$ standard deviations of these averages. The best linear fit is shown by the thin black line. {\em Bottom panel:} Mass accretion enhancement $\xi$ for different spins for simulations with $\beta_0 = 15^\circ$. Best fit lines are included separately for $a_* \le 0$ (dashed black line) and $a_* \ge 0$ (solid black line).}
\label{fig:xi}
\end{figure}

Starting with the $a_* = 0.9$ simulations, Figure \ref{fig:xi} (top panel) shows how the accretion enhancement factor $\xi$ increases with tilt $\beta_0$. Over the range of $0^\circ-30^\circ$, the dependence appears to be well fit by a linear function of the form $\xi(a_*=0.9) = 0.18\beta_0 + 0.98$ with $\chi^2 = 0.50$. Since all of the simulations in the top panel of Figure \ref{fig:xi} have the same spin, and hence the same normalization of $\xi$, the plot equivalently represents how $\dot{m}_\mathrm{BH}$ scales with $\beta_0$. Because of the design of our grid (with the polar regions being significantly less refined than the midplane), it would be difficult for us to explore tilts larger than about $\beta_0 = 30^\circ$ without significantly modifying our numerical approach. Therefore, at this time we cannot assess whether this linear trend would continue to larger tilts. 

Another trend apparent in Figure \ref{fig:mdot_tilted} is that the mass accretion rate is more variable for larger tilts. This fact is also apparent from the growing magnitude of the error bars in Figure \ref{fig:xi} (top panel) with increasing tilt. This enhanced variability of tilted discs is a subject we plan to explore further in future work.

Next we test the dependence of the accretion enhancement on spin. To do so, we consider simulations with $a_* = -0.9$, 0, 0.5, and 0.9. Since tilt is undefined for a non-spinning black hole, the accretion enhancement factor for $a_* = 0$ is 1 by definition. For the other spins, we use only the $\beta_0 = 15^\circ$ simulations. Figure \ref{fig:xi} (bottom panel) shows the dependence of $\xi$ on spin for these cases. Unsurprisingly, the mass accretion enhancement is stronger with higher spins. Again, the dependence can be fit by a linear function. For $a_* \ge 0$, we find $\xi(a_* \ge 0;\beta_0=15^\circ) = 1.67a_* + 0.876$ with $\chi^2 = 1.35$. For $a_* \le 0$, we only have two points, which can, of course, be fitted by a line. In this case, $\xi(a_*\le0;\beta_0=15^\circ) = -1.45a_* + 1.00$, which has almost the same magnitude slope as for $a_* \ge 0$. Note that in the bottom panel of Figure \ref{fig:xi}, since each simulation has a different spin, each one also has a different normalization for $\xi$. Therefore, although $\xi$ (the accretion enhancement due to tilt) increases with spin magnitude, this does not imply that $\dot{m}_\mathrm{BH}$ necessarily does as well.


\subsection{Radiative efficiency}

\begin{figure}
\centering
\includegraphics[width=1.0\linewidth,trim=0mm 0mm 0mm 0,clip]{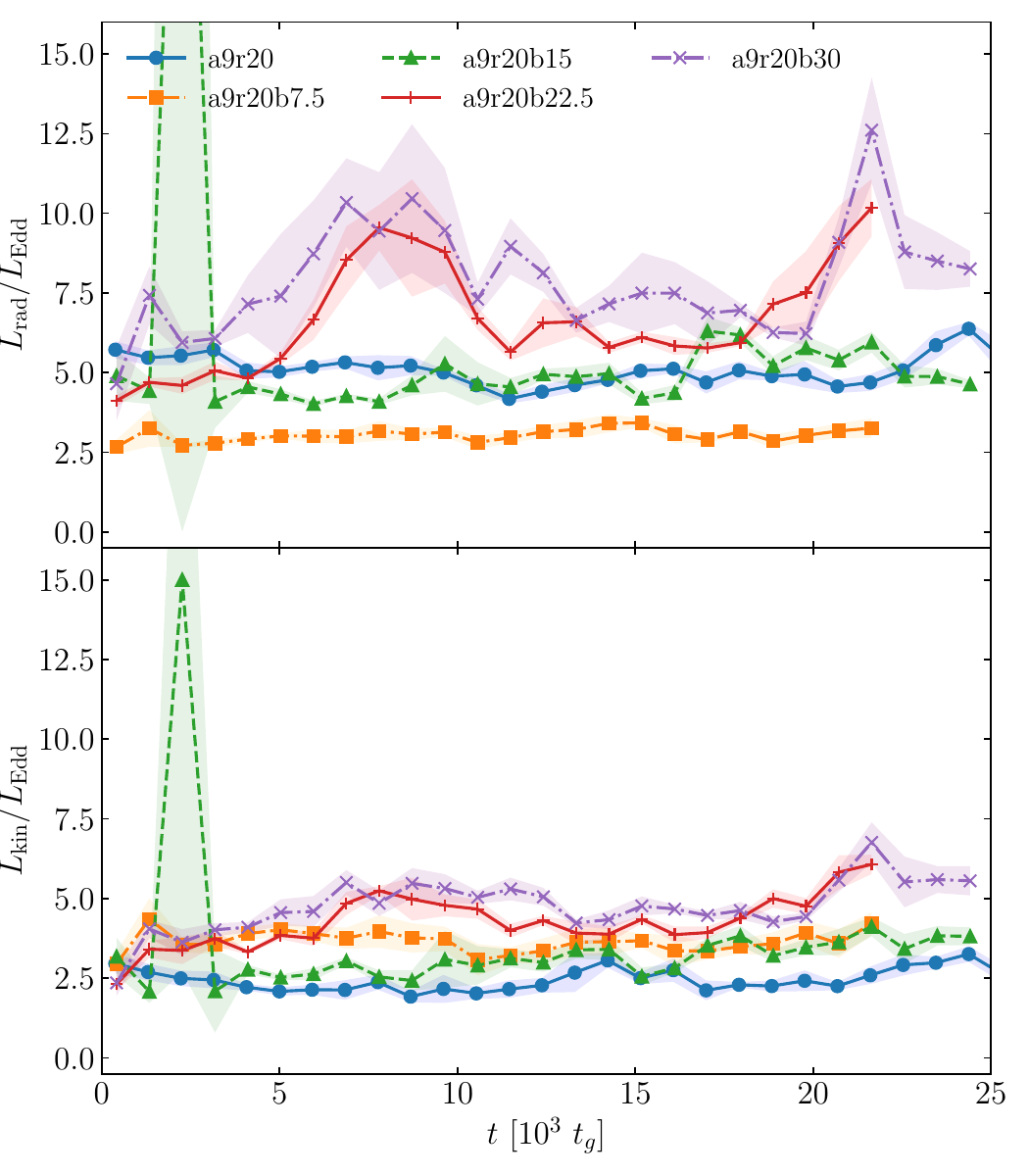}
\caption{Radiative ({\em top panel}) and kinetic ({\em bottom panel}) luminosities as a function of time, measured at the equilibrium radius $r_\mathrm{eq}$ (full $4\pi$ steradians) for the $a_* = 0.9$ simulations with tilts varying from $\beta_0 = 0^\circ$ to $30^\circ$. Data have been smoothed by using an averaging window of 20 consecutive dumps ($\approx 1850\,t_g$ in time). The shaded regions show $1\sigma$ standard deviations.}
\label{fig:luminosity}
\end{figure}

\begin{figure}
\centering
\includegraphics[width=1.0\linewidth,trim=0mm 0mm 0mm 0,clip]{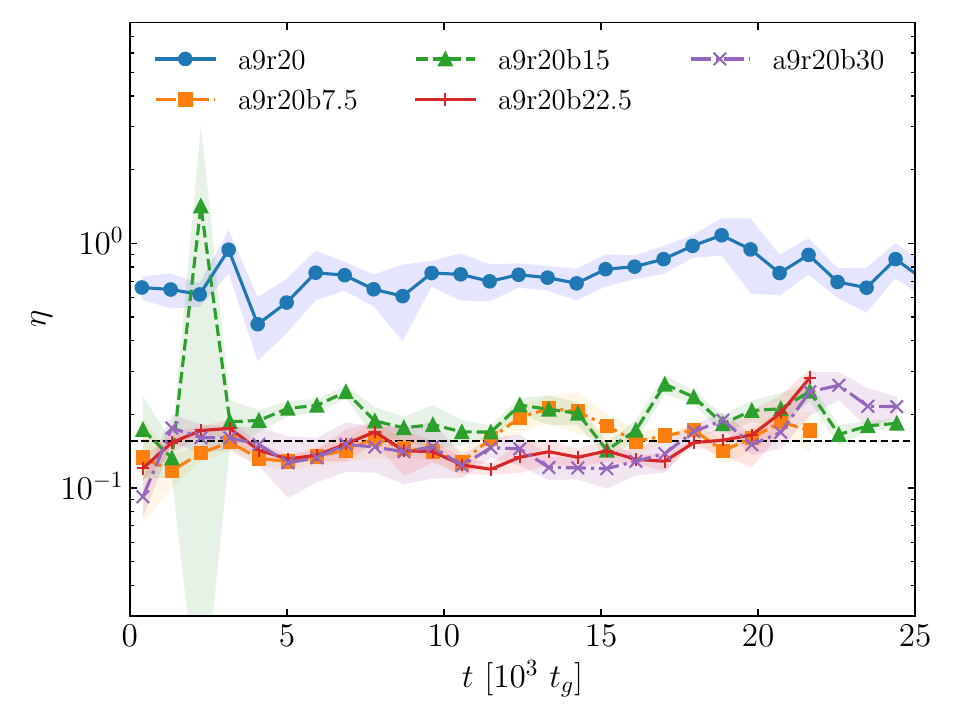}
\caption{Radiative efficiency $\eta = L/\dot{M}c^2$ as a function of time for the $a_* = 0.9$ simulations with tilts varying from $\beta_0 = 0^\circ$ to $30^\circ$. The black dashed line shows the expected efficiency for a Novikov-Thorne disc with $a_*=0.9$.}
\label{fig:efficiency}
\end{figure}

We already saw in Figures \ref{fig:mdot_tilted} and \ref{fig:xi} that tilted accretion discs have higher mass accretion rates than their untilted counterparts, exceeding the Eddington limit by up to a factor of ten (it could possibly be even higher, but our current simulation setup limits us to exploring $\beta_0 \lesssim 30^\circ$). Tilted accretion discs also appear to generally be more luminous\footnote{The $\beta_0 = 7.5^\circ$ simulation actually has a lower radiative luminosity than the untilted simulation, though it has a slightly higher kinetic luminosity.}, as shown in Figure \ref{fig:luminosity}. Each luminosity is measured at the maximum radius for which each simulation has come into inflow equilibrium, $r_\mathrm{eq}$, based on $\dot{m}_\mathrm{net}$ being flat in Figure \ref{fig:mdot_flux}. The values for $r_\mathrm{eq}$, $L_\mathrm{out}(r_\mathrm{eq})$, and $L_\mathrm{kin}(r_\mathrm{eq})$ are reported for each simulation in Table \ref{tab:models}. However, the increase in luminosity is less than the increase in mass accretion rate, so that the radiative efficiency $\eta = L/\dot{M}c^2$ of tilted discs is actually lower, as shown in Figure \ref{fig:efficiency}. Interestingly, the efficiencies of all of our $a_*=0.9$ tilted discs cluster around the value expected for a Novikov-Thorne disc with this spin, whereas the efficiency of the untilted disc is about a factor of five larger. 

We caution, however, that we consider our luminosities, and therefore our efficiencies, to be upper limits. This is because some fraction of the luminosities we report are carried in the optically thick winds, which may still be bound and ultimately fall back to the disc. Additionally, the luminosities in Figure \ref{fig:luminosity} represent integrals over the complete radial shell, so they are true, total luminosities, and are thus unlikely to match what an observer would infer from any one particular viewing angle. 

\section{Standing shocks}
\label{sec:standing_shocks}

\begin{figure}
\centering
\includegraphics[width=1.0\linewidth,trim=0mm 0mm 0mm 0,clip]{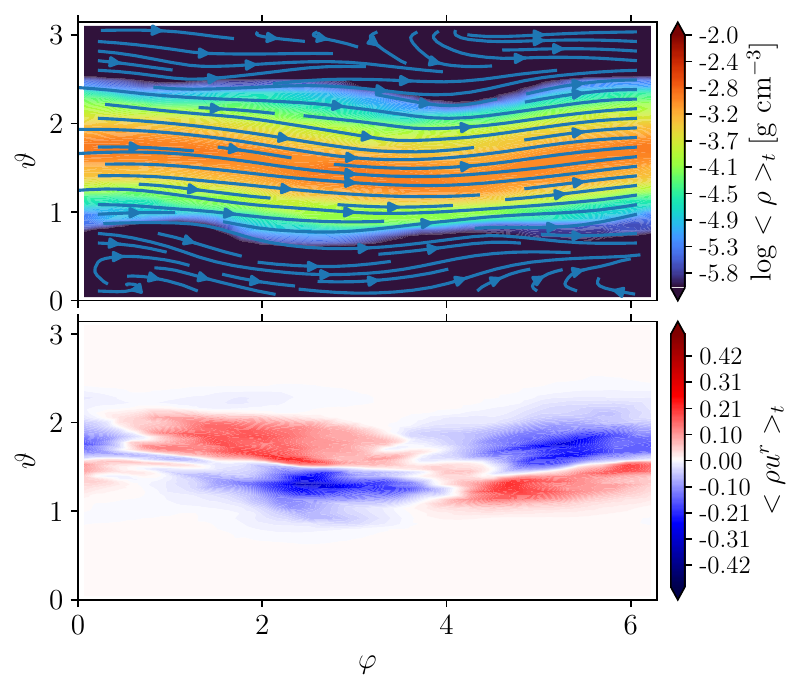}
\caption{{\it Top:} Psuedocolor plot of time-averaged gas density and fluid velocity streamlines in the grid coordinate $\{\vartheta,\varphi\}$ frame at $r=13.2\,r_g$ for simulations a9r20b15. Time averaging is over the interval from $t = 115\,000\,t_g$ to $125\,000\,t_g$. {\it Bottom:} Radial mass flux density through the same $\vartheta$-$\varphi$ slice as the top panel showing outflowing gas as red and inflowing gas as blue. The pattern is consistent with radial epicyclic (elliptical) motion, but with the top half of the disc exhibiting motion $180^\circ$ out of phase with the bottom half.}
\label{fig:slice}
\end{figure}

We now discuss the feature that truly sets tilted accretion discs apart and explains many of the differences noted in Section \ref{sec:results}. Owing to unbalanced radial pressure gradients, the gas within tilted discs is forced into radial epicyclic motion. A different way to picture this is that the particles within the disc follow elliptical trajectories. We can see evidence for these elliptical trajectories in the bottom panel of Figure \ref{fig:slice}, which shows the radial mass flux density over a shell of constant radius ($r=13.2\,r_g$), with red colors showing matter moving outward (toward larger radii) and blue colors showing matter moving inward (toward smaller radii). Note that this is collective behavior. It is not that every particle is following a random elliptical trajectory, but rather that an organized, systemic motion is set up in the discs.

\begin{figure}
\centering
\includegraphics[width=1.0\linewidth,trim=0mm 0mm 0mm 0,clip]{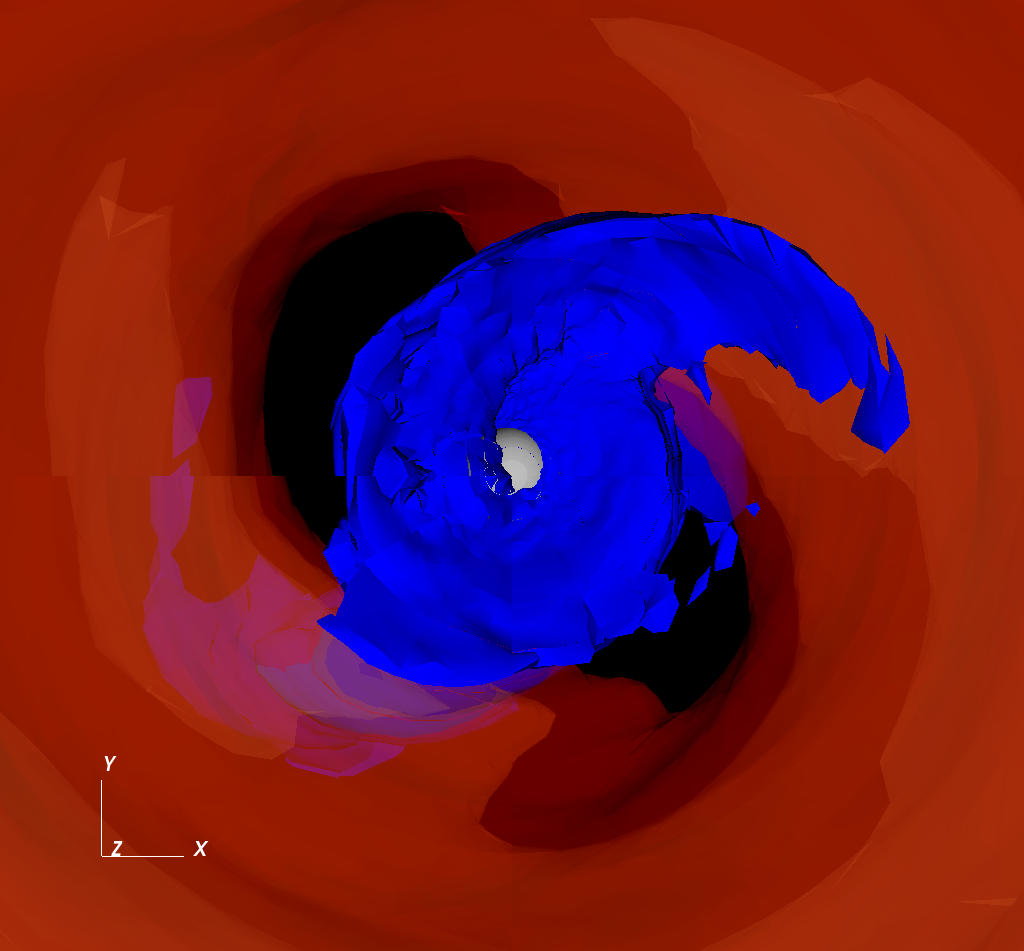}
\caption{Isosurface plots of density (semitransparent red) and $\vert \mathrm{curl}\,\mathbf{V}\vert$ (blue) for simulation a9r20b15. The quantity $\vert \mathrm{curl}\,\mathbf{V}\vert$ is a good tracer of the location of a shock. The plot is restricted so that we only show the shock surface in regions where $\rho \ge 10^{-5}$ g cm$^{-3}$ to prevent overcrowding of the image from shocks associated with the lower-density outflowing winds. This figure is oriented looking directly down the initial symmetry axis of the accretion disc. The spin axis of the hole is tilted $15^\circ$ to the left in the image. Note that one part of the shock surface is found above the midplane of the disc (upper right part of image), while another part is located below the midplane (partially hidden in the lower left part of the image).}
\label{fig:shock}
\end{figure}

There are two important points to make about these elliptical trajectories: 1) they are $180^\circ$ out of phase across the midplane of the disc \citep[as previously noted in][]{Fragile08}; and 2) their eccentricities increase with decreasing radius \citep[previously noted in][]{Dexter11}. A consequence of the second point is that there is a crowding of trajectories near their respective apocenters (denoted by $r_a$). As a way to understand this, imagine one particle with a large orbit (large semi-major axis $a$) but small eccentricity $e$ and another particle with a small orbit but a large eccentricity. Since $r_a = a(1+e)$, these two particles can have their apocenters located at the same radius even though the orbits have different sizes. Whenever this happens, these particles have the potential to collide. In a disc, there are many such particles experiencing this crowding of orbits. This leads to a significant compression (or shock) forming within the disc. However, because of the first point about the elliptical trajectories being $180^\circ$ out of phase across the disc midplane, there are actually two shocks, one for the gas in the upper half of the disc that forms on one side of the black hole and another for the gas in the lower half that forms on the opposite side. This pattern of two opposing standing shocks is illustrated in Figure \ref{fig:shock} for simulation a9r20b15. These shocks are stable features whose locations remain fixed relative to the line of nodes between the disc midplane and black hole symmetry plane, thus precessing with the disc \citep{Fragile08}.

Although these shocks extend only a few $r_g$ in radius, they can have a profound effect on the dynamics. They lead to significant dissipation \citep{Generozov14}, which increases the entropy and temperature of the gas \citep{Fragile08, Dexter11}. They also extract angular momentum \citep{Fragile07, Fragile08}, which explains the enhanced mass accretion rate noted in Section \ref{sec:mdot_tilted}. These standing shocks are also possible sites of particle acceleration \citep{Sironi24}.

Shocks have been noted in other tilted accretion disc simulations \citep{Kaaz23, Liska23}. However, the standing shocks highlighted here appear to be distinct from the ``nozzle'' shocks presented in those works. For one thing, the standing shocks are found out of (both above and below) the disc midplane, whereas the nozzle shocks correspond to vertical motion converging at the midplane. Furthermore, we see no evidence in our simulations for the vertical bouncing and compression that are the root causes of the nozzle shocks (see the top panel of Figure \ref{fig:slice} and compare to Figure 2 of \citet{Kaaz23}). Finally, our standing shocks are only found relatively close to the black hole, whereas the nozzle shocks of \citet{Kaaz23} extend along the entire length of the line-of-nodes created by the disc midplane and black hole symmetry plane.

\section{Lense-Thirring Precession}
\label{sec:precession}

\begin{figure*}
\centering
\includegraphics[width=1.0\textwidth,trim=0mm 0mm 0mm 0,clip]{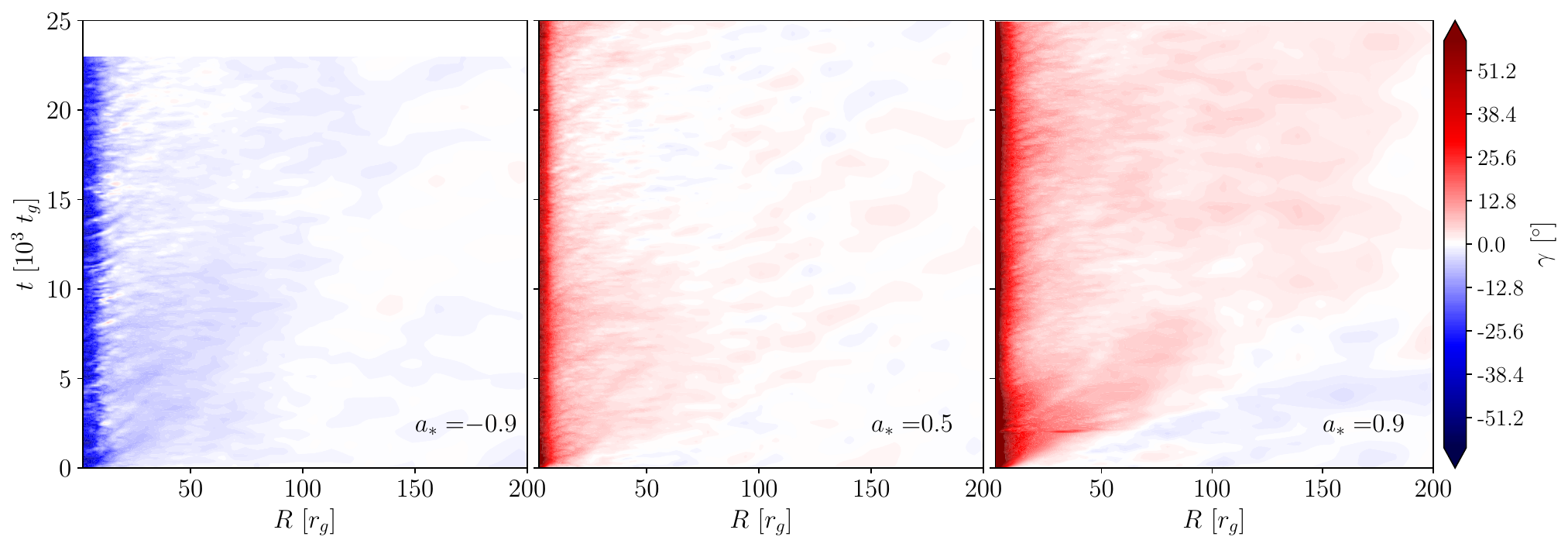}
\caption{Spacetime plots of the twist or precession angle, $\gamma$ (measured in degrees), for the $\beta_0 = 15^\circ$ tilted simulations a-9r200b15 (left), a5r50b15 (middle) and a9r20b15 (right). There is no evidence of significant, sustained global precession over any region.}
\label{fig:twist}
\end{figure*}

As mentioned in Section \ref{sec:introduction}, a prominent characteristic of tilted accretion discs is Lense-Thirring precession \citep[see][and references therein]{Fragile25b}. Two possible outcomes of this precession are Bardeen-Petterson alignment or global, solid-body precession. In the only previous work to consider tilted supercritical accretion discs, \citet{Asahina24} found that their discs underwent global precession on reasonably short timescales (estimated to be about 10 s for a $10 M_\odot$ black hole). However, while those simulations shared many similarities with our own in terms of numerical methods -- GRRMHD with $\mathbf{M}_1$ closure on a spherical-polar grid -- they differed in one critical aspect: while \citet{Asahina24} considered a gas torus with a small radial extent (only a few tens of $r_g$), we simulated very large Keplerian discs. As a consequence, our discs are much too large to globally precess on the timescale of even our longest duration simulations. This lack of significant global precession can be confirmed in the spacetime diagrams in Figure \ref{fig:twist}. Other than a small region inside $10\,r_g$ that undergoes rapid differential precession and then freezes, the rest of the disc only precesses a few degrees before seeming to stall. Nevertheless, Figure \ref{fig:twist} does confirm one of our expectations. Our positive spin cases (a5r50b15 and a9r20b15) show prograde precession, while our negative spin one (a-9r200b15) precesses in a retrograde fashion.

\begin{figure*}
\centering
\includegraphics[width=1.0\textwidth,trim=0mm 0mm 0mm 0,clip]{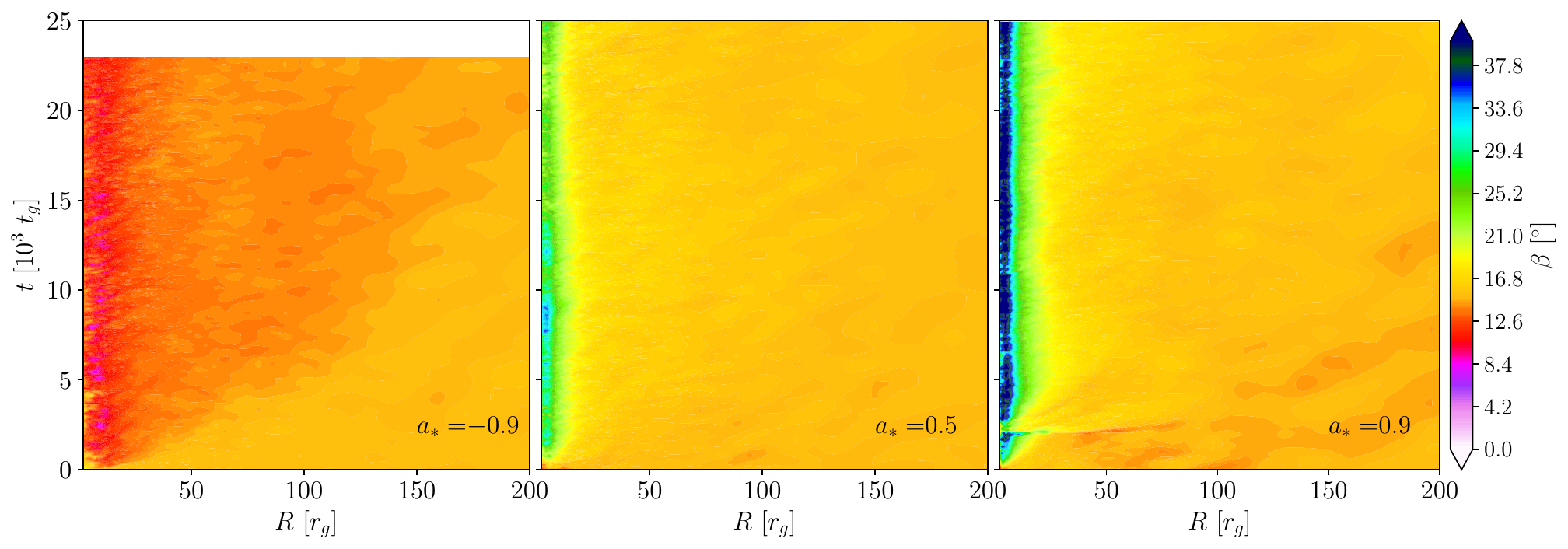}
\caption{Spacetime plots of the tilt, $\beta$ (measured in degrees), for the simulations with initial tilts of $15^\circ$: a-9r200b15 (left), a5r50b15 (middle) and a9r20b15 (right). Most of the disc remains tilted at $\approx 15^\circ$. However, the prograde discs have a local peak in $\beta$ at small radii, while the retrograde case exhibits lower values of $\beta$ at small radii, which in this case means the disc is bending toward full anti-alignment.}
\label{fig:tilt}
\end{figure*}

There is similarly little evidence for significant Bardeen-Petterson alignment. In fact, as shown in Figure \ref{fig:tilt}, the prograde cases (a5r50b15 and a9r20b15) exhibit tilt profiles that actually increase at small radii rather than decrease. This is consistent with the findings of previous (non-radiative) tilted disc simulations \citep{Fragile07, Liska18, White19}. The retrograde case (a-9r200b15), on the other hand, shows evidence of modest alignment (the tilt drops from $15^\circ$ to about $10^\circ$). A similar tendency for prograde discs to bend away from alignment and retrograde discs to bend toward anti-alignment was noted in \citet{Morales14}. 

To summarize, our discs all maintain tilted, warped configurations with the Lense-Thirring precession torque being transmitted slowly throughout the entire disc and any significant precession happening on timescales too long to measure in these simulations. It is also noteworthy that our discs do not tear or otherwise separate into regions that precess on different timescales. 


\section{Conclusions}
\label{sec:conclusions}

This first detailed study of tilted supercritical accretion discs has yielded a number of groundbreaking discoveries. First, while our untilted discs closely adhere to the Eddington limit\footnote{Other works, such as \citet{Yoshioka22, Zhang25}, suggest that the Eddington limit can be exceeded, even for large, Keplerian discs, provided enough material is available in the disc. This discrepancy is a topic that requires further investigation in future work.} (see Figure \ref{fig:mdot_untilted}), the tilted ones can exceed it by at least a factor of a few (Figure \ref{fig:mdot_tilted}), provided the black hole is spinning moderately rapidly ($a_* \gtrsim 0.5$) and that the tilt is nontrivial ($\beta_0 \gtrsim 10^\circ$). The simulations considered in this work suggests that the mass accretion enhancement $\xi = \dot{m}_\mathrm{BH}(\beta_0)/\dot{m}_\mathrm{BH}(0^\circ)$ may depend linearly on both quantities, i.e., $\xi \propto \beta_0$ for a fixed $a_*$ and $\xi \propto \vert a_* \vert$ for a fixed $\beta_0$ (see Figure \ref{fig:xi}).

These findings could be significant as they offer a simple solution to the problem of growing the first supermassive black holes in the Universe (Middleton et al., in prep). Because of the exponential nature of the growth, even a relatively modest enhancement above $\dot{M}_\mathrm{Edd}$ can be enough to reach the required masses in the available time. This solution is also consistent since SMBHs can be expected to repeatedly accrete from tilted discs, as described in the Introduction.

In line with the finding of more efficient mass accretion, we also discovered that tilted supercritical discs are more advective than their untilted counterparts. This is supported by a number of lines of evidence. For instance, in Figure \ref{fig:mdot_flux}, the mass outflow component $\dot{m}_\mathrm{out}$, which at most radii in these supercritical discs has a magnitude comparable to $\dot{m}_\mathrm{in}$, remains zero over a larger range of radii close to the black hole whenever the tilt increases. It logically follows that, if the outflow only becomes non-zero further out, then the inner regions of these flows must be more dominated by inflow. This is also seen in how the location of the trapping radius $r_\mathrm{tr}$ moves further out for tilted discs (see Figure \ref{fig:Lrad_flux}). 

Along with higher mass accretion rates, our tilted discs also generally produce higher luminosities (Figure \ref{fig:luminosity}). However, the increase in luminosity is less than the increase in mass accretion rate, making our tilted discs less radiatively efficient than their untilted counterparts (Figure \ref{fig:efficiency}).

Another discovery is that the mass accretion rate and luminosity of these supercritical discs become more variable as their tilt increases. This is seen in Figures \ref{fig:mdot_tilted} and \ref{fig:luminosity} and in the error bars in Figure \ref{fig:xi} (top panel). The association of variability with tilt has gained appreciation recently \citep[see][]{Fragile25b}, and although we have yet to find any strong quasi-periodic behavior in these simulations, this is something we plan to investigate more thoroughly in the future.

Most of the unique behaviors of these tilted supercritical discs can be attributed to the two opposing standing shocks that form in them. These shocks are the result of latitude-dependent radial epicyclic motion, triggered by unbalanced radial pressure gradients introduced by the tilt. Because the eccentricity of the orbital trajectories increases with decreasing radius, there is a crowding of orbits near their respective apocenters. Close to the black hole, this crowding becomes strong enough to manifest as shocks. Because the epicyclic motion is $180^\circ$ out of phase across the disc midplane, the two shocks form on opposite sides of the black hole, one above the disc midplane and the other below (as seen in Figure \ref{fig:shock}).

Finally, we found that there is no significant precession in these simulations (Figure \ref{fig:twist}). These discs are too large to precess appreciably even on the rather long timescales of these simulations. Furthermore, even our most tilted discs show no sign of tearing or otherwise separating into regions that might be able to precess on different timescales. There is also no evidence for Bardeen-Petterson alignment in our prograde tilted discs and only a limited tendency toward anti-alignment in our retrograde case (Figure \ref{fig:tilt}).

\section*{Acknowledgements}

We would like to thank the anonymous referee for their help in improving this manuscript. PCF gratefully acknowledges the support of NASA under award No 80NSSC24K0900. MJM gratefully acknowledges the support of STFC (ST/Y001699/1). The Flatiron Institute is a division of the Simons Foundation. Resources supporting this work were provided by the NASA High-End Computing (HEC) Program through the NASA Advanced Supercomputing (NAS) Division at Ames Research Center. This work used the DiRAC Memory Intensive service (Cosma8) at Durham University, managed by the Institute for Computational Cosmology on behalf of the STFC DiRAC HPC Facility (www.dirac.ac.uk). The DiRAC service at Durham was funded by BEIS, UKRI and STFC capital funding, Durham University and STFC operations grants. DiRAC is part of the UKRI Digital Research Infrastructure. The authors acknowledge the use of resources provided by the Isambard 3 Tier-2 HPC Facility. Isambard 3 is hosted by the University of Bristol and operated by the GW4 Alliance (https://gw4.ac.uk) and is funded by UK Research and Innovation; and the Engineering and Physical Sciences Research Council [EP/X039137/1].

\section*{Data Availability}

The data underlying this paper will be shared upon reasonable request to the corresponding author.



\bibliographystyle{mnras}







\bsp	
\label{lastpage}
\end{document}